\pdfoutput=1
\documentclass[preprint,1p,times,twocolumn,longmaketitle]{elsarticle}

\usepackage{amssymb}
\usepackage{amsmath}

\usepackage[numbers]{natbib}
\usepackage{xcolor}
\usepackage{hyperref}

 \usepackage{lineno}

\journal{Physics Letters B}

\begin{document}

\begin{frontmatter}

\title{Measurement of three-dimensional inclusive muon-neutrino charged-current cross sections on argon with the MicroBooNE detector}

\author[Tufts]{P.~Abratenko}
\author[Tufts]{O.~Alterkait}
\author[IIT]{D.~Andrade~Aldana}
\author[Manchester]{L.~Arellano}
\author[UTA]{J.~Asaadi}
\author[TelAviv]{A.~Ashkenazi}
\author[FNAL]{S.~Balasubramanian}
\author[FNAL]{B.~Baller}
\author[Oxford]{G.~Barr}
\author[Oxford]{D.~Barrow}
% ...
\cortext[eee]{microboone\_info@fnal.gov}

% ...
\address[FNAL]{Fermi National Accelerator Laboratory (FNAL), Batavia, IL 60510, USA}
% ...
\address[IIT]{Illinois Institute of Technology (IIT), Chicago, IL 60616, USA}
% ...
\address[Manchester]{The University of Manchester, Manchester M13 9PL, United Kingdom}
% ...
\address[Oxford]{University of Oxford, Oxford OX1 3RH, United Kingdom}
% ...
\address[TelAviv]{Tel Aviv University, Tel Aviv, Israel, 69978}
% ...
\address[UTA]{University of Texas, Arlington, TX, 76019, USA}
\address[Tufts]{Tufts University, Medford, MA, 02155, USA}
% ...

\begin{abstract}
We report the measurement of the triple-differential cross section $d^{3}\sigma / dE_{\mathrm{vis}} d\cos(\theta_{\mu}) dP_{\mu}$ for inclusive muon-neutrino charged-current scattering on argon.  This measurement utilizes data from 6.4$\times10^{20}$ protons on target of exposure collected using the MicroBooNE liquid argon time projection chamber located along the Fermilab Booster Neutrino Beam with a mean neutrino energy of approximately 0.8~GeV. The mapping from reconstructed kinematics to truth quantities is validated within uncertainties by comparing the distribution of reconstructed hadronic energy in data to that of the model prediction in different muon scattering angle bins after applying a conditional constraint from the muon momentum distribution in data. The success of this validation provides confidence that the energy transfer in the MicroBooNE detector is well-modeled within simulation uncertainties, enabling a reliable unfolding to a triple-differential cross section defined at the nominal neutrino flux over muon momentum, muon scattering angle, and visible neutrino energy. This validation not only supports accurate cross-section extraction, but also establishes a critical foundation for tuning interaction models used in future neutrino oscillation measurements. The unfolded measurement covers an extensive phase space, providing a wealth of information useful for future liquid argon time projection chamber experiments measuring neutrino oscillations.  Comparisons against a number of commonly used model predictions are included and their performance in different parts of the available phase-space is discussed.
\end{abstract}

\begin{keyword}
neutrino experiment \sep neutrino cross-section \sep charged-current interactions



\end{keyword}

\end{frontmatter}



Precision modeling of neutrino-nucleus interactions is necessary to achieve the goals of future accelerator neutrino oscillation experiments.  Neutrino cross-section modeling is one of the dominant sources of uncertainty in the current generation of oscillation experiments~\cite{PhysRevD.106.032004,t2k_osc} and could in principle limit the search for leptonic charge-parity violation~\cite{NAGU2020114888,hyperk2025,ALVAREZRUSO20181}.  In the energy range of 0.1--5\,GeV, the dominant modes of neutrino interactions, such as quasi-elastic (QE) scattering and resonance production, are difficult to model because of various nuclear effects.  Typical examples include nuclear ground state modeling, nucleon-nucleon correlations, and final state interactions~\cite{RevModPhys.84.1307,NUSTEC2025}.  Efforts to simulate these interactions accurately would benefit from dedicated measurements that probe the combined phase space of leptonic and hadronic kinematics.  For inclusive muon neutrino ($\nu_{\mu}$) charged current (CC) scattering, there are three degrees of freedom determining the principle interaction kinematics: the scattering muon momentum ($P_{\mu}$) and angle ($\theta_{\mu}$) that are directly measured, and the neutrino energy ($E_{\nu}$) that is deduced with the measurement of the hadronic energy. The accurate reconstruction of the neutrino energy is of particular importance to upcoming precision long-baseline neutrino oscillation measurements~\cite{Abi_2020,Antonello:2015lea}.

There have been continuous advancements in the field of inclusive and exclusive neutrino-nucleus scattering (see Ref.~\cite{PhysRevD.110.092014,PhysRevLett.133.041801,minerva_ma,PhysRevD.109.092008,qh28-4znk,PhysRevD.108.112009}
among others for recent progress). Of particular interest to the measurement presented in this article is a recent triple-differential cross section measured on carbon at MINER$\nu$A, where the independent variables are the muon kinematics and the total observed proton energy~\cite{MINERvA:2022mnw}. On an argon target, single- and double-differential $\nu_{\mu}$ CC inclusive cross sections have been reported~\cite{PhysRevLett.108.161802,Acciarri_2014,MCC8_2D_xs,wc_1d_xs}.  The measurement presented here expands upon the work measuring energy-dependent cross sections in Ref.~\cite{wc_1d_xs}.  Specifically, we report the first measurement of the nominal-flux-averaged inclusive $\nu_{\mu}$ CC triple-differential cross section on argon $d^{3}\sigma /dE_{\mathrm{vis}} d\cos(\theta_{\mu}) dP_{\mu}$. The visible neutrino energy, $E_{\mathrm{vis}}$, is defined as the truth level sum of reconstructable energy deposited, which includes all final state particles, besides neutrons, with no thresholds. Particle masses are included, except for protons which are assigned a binding energy of 8.6$\,$MeV.  Neutrino events are selected using the $\nu_{\mu}$ selection described in~\cite{wc_elee_prd}, with $E_{\nu} \in [0.2,4.0]$\,GeV and $P_{\mu} \in [0,2.5]$\,GeV/$c$, giving an overall selection efficiency of 68\% and purity of 92\%. The same event selection is used to measure the three-dimensional cross section $d^{2}\sigma(E_{\nu}) / d\cos(\theta_{\mu}) dP_{\mu}$, replacing $E_{\mathrm{vis}}$ with $E_{\nu}$, which is used to demonstrate the success of the model validation. 
The estimation of the visible neutrino energy uses the measurements of the reconstructed visible hadronic energy, $E^{\mathrm{rec}}_{\mathrm{had}}$, and reconstructed muon momentum, $P^{\mathrm{rec}}_{\mu}$. Therefore, these quantities are investigated before being combined to form $E^{\mathrm{rec}}_{\mathrm{vis}}$ in our model validation tests. Leveraging high statistics and extensive three-dimensional phase-space coverage, we extend the validation procedure~\cite{MicroBooNE:2024kwe} initially introduced in~\cite{wc_1d_xs} to probe the modeling of kinematics in multiple dimensions. This method employs $\chi^{2}$ goodness-of-fit tests to compare reconstructed data distributions with model predictions, ensuring that model uncertainties adequately encompass the observed discrepancies. Additionally, by incorporating muon kinematics as a constraint, the procedure enhances sensitivity to the modeling of missing hadronic energy, providing a more stringent and reliable validation.

The MicroBooNE liquid argon time projection chamber (LArTPC) measures 2.56\,m along the drift direction, 10.36\,m along the beam direction, and 2.32\,m along the vertical direction.  It has an active mass of 85 tonnes of LAr and is capable of mm-level position resolution, as well as calorimetry with MeV-level detection threshold~\cite{uboone_detector}.  Ionization electrons drift in a 273\,V/cm electric field towards an anode consisting of 3 detection planes of wires at $60^\circ$ angles to each other with a wire pitch of 3\,mm.  Thirty-two photomultiplier tubes (PMTs) are used to detect the scintillation light from the interaction to provide a prompt timing signal.  The Booster Neutrino Beam (BNB) at Fermilab produces neutrinos at a target 470\,m upstream of the MicroBooNE detector, with 93.6\% estimated to be $\nu_{\mu}$ at a mean $E_{\nu}$ of 0.8\,GeV~\cite{AguilarArevalo:2008yp}.

The event selection used in this analysis is the same as the $\nu_{\mu}$ CC selection used in the MicroBooNE inclusive $\nu_{e}$ low-energy excess search~\cite{wc_elee_prd}, and was performed on a data set collected from 2016--2018 using an exposure of $6.4 \times 10^{20}$ protons on target (POT), an order of magnitude larger than the single-differential energy-dependent cross section measurement presented in~\cite{wc_1d_xs}.  The Wire-Cell reconstruction chain leverages the detector information through the use of tomography, matching of TPC-charge clusters to PMT-light flashes, and trajectory fitting for particle identification and cosmic-ray removal~\cite{Qian:2018qbv,MicroBooNE:2021zul}.  Higher-level algorithms perform pattern recognition, neutrino vertex identification, topology classification, and particle identification to produce a particle flow within an event~\cite{wc_pattern_recognition}.  The boosted-decision-tree library XGBoost~\cite{Chen:2016btl} is then used to further reduce backgrounds to achieve the $\nu_{\mu}$ CC selection.

Energy reconstruction is crucial for the extraction of energy-dependent cross sections~\cite{wc_1d_xs} as well as for the search for new physics beyond the Standard Model~\cite{wc_elee_prd}.  Generally, energy reconstruction is separated into the reconstruction of particle tracks and of electromagnetic (EM) showers.  By default, particle tracks have their energy estimated from their propagation length using a tabulation of the Bethe-Bloch formula from the NIST PSTAR database~\cite{pstar}.  This method is substituted with a calorimetry-based approach in cases where the range-based estimation is poor, including short tracks ($<$ 4\,cm), tracks exiting the detector, tracks that frequently change directions, and muon tracks with identified $\delta$ rays~\cite{wc_pattern_recognition}.  The calorimetry-based approach uses a recombination model~\cite{Adams:2019ssg} to convert the measured $dQ/dx$ to the energy loss per unit length $dE/dx$, which is then integrated.  The estimation of EM shower energy also follows a calorimetry-based approach, but uses the total measured charge and a different scale factor~\cite{MicroBooNE:2020vry} that includes the overall mean recombination effect as well as contributions for clustering efficiency and detection threshold.  This scaling factor is validated through the reconstructed invariant mass of the neutral pion~\cite{Adams:2019law}.  Reconstructed muons, charged pions, and electron candidates have their mass added to their energy reconstruction, and proton candidates are assigned an average binding energy of 8.6\,MeV~\cite{Sukhoruchkin:amnbe}.  The reconstructed visible neutrino energy is constructed as the sum of the muon and hadronic energies.  Energy resolutions are estimated from Monte Carlo simulation~\cite{genie-tune-paper}.  For $\nu_{\mu}$ CC events with their main TPC cluster fully contained within the fiducial volume (fully contained events), the estimated kinematic resolutions are $\approx$10\% on muon energy, $\approx$30--50\% on energy transfer, defined as $E_{\nu} - E_{\mu}$, resulting from imperfect reconstruction and missing hadronic energy $E^{\mathrm{missing}}_{\mathrm{had}}$, and $\approx$20\% on $E_{\mathrm{vis}}$.  The angular resolution reaches $5^{\circ}$ in $\theta_{\mu}$ at forward angles, but is less accurate at backwards angles.

The neutrino flux prediction is derived from the MiniBooNE flux simulation~\cite{AguilarArevalo:2008yp} updated to the MicroBooNE detector location, with muon neutrino flux prediction uncertainties ranging from 5--15\% over the flux range of $\approx$$0.1$--$4.0\,$GeV.  Neutrino-argon ($\nu$-Ar) interactions are modeled using \texttt{GENIE} v3.0.6 G18\_10a\_02\_11a tuned to T2K data~\cite{genie-tune-paper,T2K:2016jor}, referred to as the \texttt{MicroBooNE model}.  In particular, hadronic interactions contributing to missing energy are conservatively estimated, with proton-to-neutron conversion and proton knockout having 50\% and 20\% uncertainties respectively~\cite{Andreopoulos:2015wxa}. The model also includes a conservative 50\% uncertainty on the 2p2h normalization.  Overall, there is a $\approx$20\% $\nu$-Ar interaction uncertainty on the measurement.  Measurement uncertainties on flux, cross section, and secondary interactions of protons and charged pions outside the target nucleus (0.6\% of the reported cross section, simulated with \texttt{GEANT4}~\cite{geant_4}) are each modeled using a multisim technique to calculate a covariance matrix~\cite{unisim}.  Additionally, uncertainties are included for the model simulation statistics that are estimated using the Poisson likelihood method~\cite{Arguelles:2019izp} ($10\%$), the modeling of ``dirt" events originating outside the cryostat~\cite{wc_elee_prd} (below $1\%$), the POT (2\%) based on measurements of the originating proton flux~\cite{AguilarArevalo:2008yp}, and the number of target nuclei (1\%).  Additional plots showing the breakdown of total uncertainties by type and the total fractional uncertainties can be found in the supplemental material~\cite{suppl}.

The detector response uncertainty considers the same effects as in previous work~\cite{det_unc,wc_1d_xs} and takes into account the impact of variations in TPC waveform, light yield and propagation, the space charge effect, and ionization recombination~\cite{Abratenko:2020bbx,Adams:2019qrr,Adams:2019ssg}.  A fixed set of MC interactions are simulated multiple times, first using parameter central values (CV) and then individually applying a $1\sigma$ variation to each parameter.  To compensate for limited simulation statistics, re-sampling of events is performed through a bootstrapping procedure, as discussed in Ref.~\cite{wc_elee_prd}.  For each parameter, we compute the average difference vector between the CV and $1\sigma$ variation, $\vec{V}_{D}^{\mathrm{nominal}}$, as well as its estimated covariance $M_{R}$.  These are used to construct a normal distribution of variations that is repeatedly sampled in the formation of the overall detector response covariance matrix $M_{D}$.

Because simulating events and propagating them through the detector is computationally expensive, there is a limited quantity of simulated events available.  The large number of bins involved in a three-dimensional analysis leads to a small number of events per bin, causing large statistical fluctuations in $\vec{V}_{D}^{\mathrm{nominal}}$, and an over-estimation of the covariance in $M_{R}$ and subsequently $M_{D}$.  Furthermore, measured correlations between bins cause the over-estimated covariance $M_{R}$ to impact all bins in the measurement, not just those with large statistical fluctuations.  To address this, a Gaussian Processes Regression (GPR) smoothing algorithm~\cite{gpr_textbook, gpr_paper, PhysRevD.101.012001} is applied to the distribution in $\vec{V}_{D}^{\mathrm{nominal}}$, smoothing the statistical fluctuations introduced by the bootstrapping procedure.  GPR uses a Bayesian approach to model the data with a joint Gaussian distribution and an uninformed prior.  A smoothed posterior is computed from the simulated values of $\vec{V}_{D}^{\mathrm{nominal}}$, as well as a kernel matrix $\Sigma_{K}$ that asserts our intuition of smoothness between nearby bin centers $x_{1}$, $x_{2}$ through a radial basis kernel function $K(x_1,x_2) = e^{-|(\vec{x}_1-\vec{x}_2) \cdot \vec{s}|^2/2}$.  Based on reconstruction resolutions~\cite{wc_pattern_recognition}, length scales $L_{i}$ were chosen to be 0.1 in $\cos(\theta^{\mathrm{rec}}_{\mu})$ and 20\% for each of $E_{\mathrm{vis}}^{\mathrm{rec}}$ and $P^{\mathrm{rec}}_{\mu}$ to calculate $s_i = 1/L_i$.  The supplemental material~\cite{suppl} provides additional details on the implementation of GPR smoothing in this work.  The central value and covariance of the posterior prediction are used in place of the original $\vec{V}_{D}^{\mathrm{nominal}}$ and $M_{R}$.  Because of GPR smoothing, statistical fluctuations are controlled and become less impactful in $M_{D}$, reducing the overall detector systematics covariance by an order of magnitude to $\approx$20\%.  The validity of this reduction is tested through the data/simulation goodness-of-fit (GoF) tests.

Since the \texttt{MicroBooNE model} is used to estimate the selection efficiency and unfold the reconstructed variables, such as $E_{\mathrm{vis}}^{\mathrm{rec}}$, to truth quantities, it is important to validate its accuracy. Furthermore, this mapping is important to compare with model predictions, which can only be made in the nominal flux. If this model (including its uncertainties) is unable to describe the distribution in data, it may introduce significant bias beyond the uncertainties into the extracted cross sections. Therefore, a comprehensive set of data/simulation comparisons using the reconstructed kinematic variables $P^{\mathrm{rec}}_{\mu}$, $\cos(\theta^{\mathrm{rec}}_{\mu})$, and $E_{\mathrm{had}}^{\mathrm{rec}}$ are investigated and discussed below, demonstrating the validity of the model.  Since $E_{\mathrm{had}}^{\mathrm{rec}}$ represents the reconstruction of the visible component of the energy transfer, it is the ideal distribution to study in complement with the muon kinematics, which together account for $E_{\mathrm{vis}}^{\mathrm{rec}}$.

Because theoretical predictions are defined under the nominal flux assumption, measured cross sections must be translated from the real flux exposure to the nominal one, which can also serve as a common basis for model comparisons. Therefore, the mapping from reconstructed to true $E_{\nu}$ needs special attention. Model dependence from this mapping is inherent to neutrino cross-section measurements, whether analyzed as a function of neutrino energy or visible quantities~\cite{MicroBooNE:2024kwe}, although the level of dependence varies between measurements. For neutrino energy, this dependence emerges during the unfolding from reconstructed to true energy, whereas for muon kinematics, it arises from mapping cross-sections to the nominal neutrino spectrum — an essential basis for comparing measurements to model predictions. This mapping is tested through the combination of GoF tests over the muon kinematics and GoF tests over $E_{\mathrm{had}}^{\mathrm{rec}}$, and the model is only considered to be validated if all tests show $\chi^{2}/\textrm{ndf}$ consistent within a $2\sigma$ level of agreement.  These first tests investigate the modeling of the muon kinematics as a prerequisite for their use as a constraint on the $E_{\mathrm{had}}^{\mathrm{rec}}$ prediction, and are performed over the two-dimensional (2D) $\{ P^{\mathrm{rec}}_{\mu}, \cos(\theta^{\mathrm{rec}}_{\mu}) \}$ distribution and are shown in the supplemental material~\cite{suppl}.  They give $\chi^{2}/\textrm{ndf}$ of 105/144 and 103/144 for the fully and partially contained events respectively, demonstrating that the model is able to describe the distribution of muon kinematics seen in data well within the model uncertainties.

\begin{figure}[hbt!]
     \centering
     \includegraphics[clip,width=0.95\textwidth]{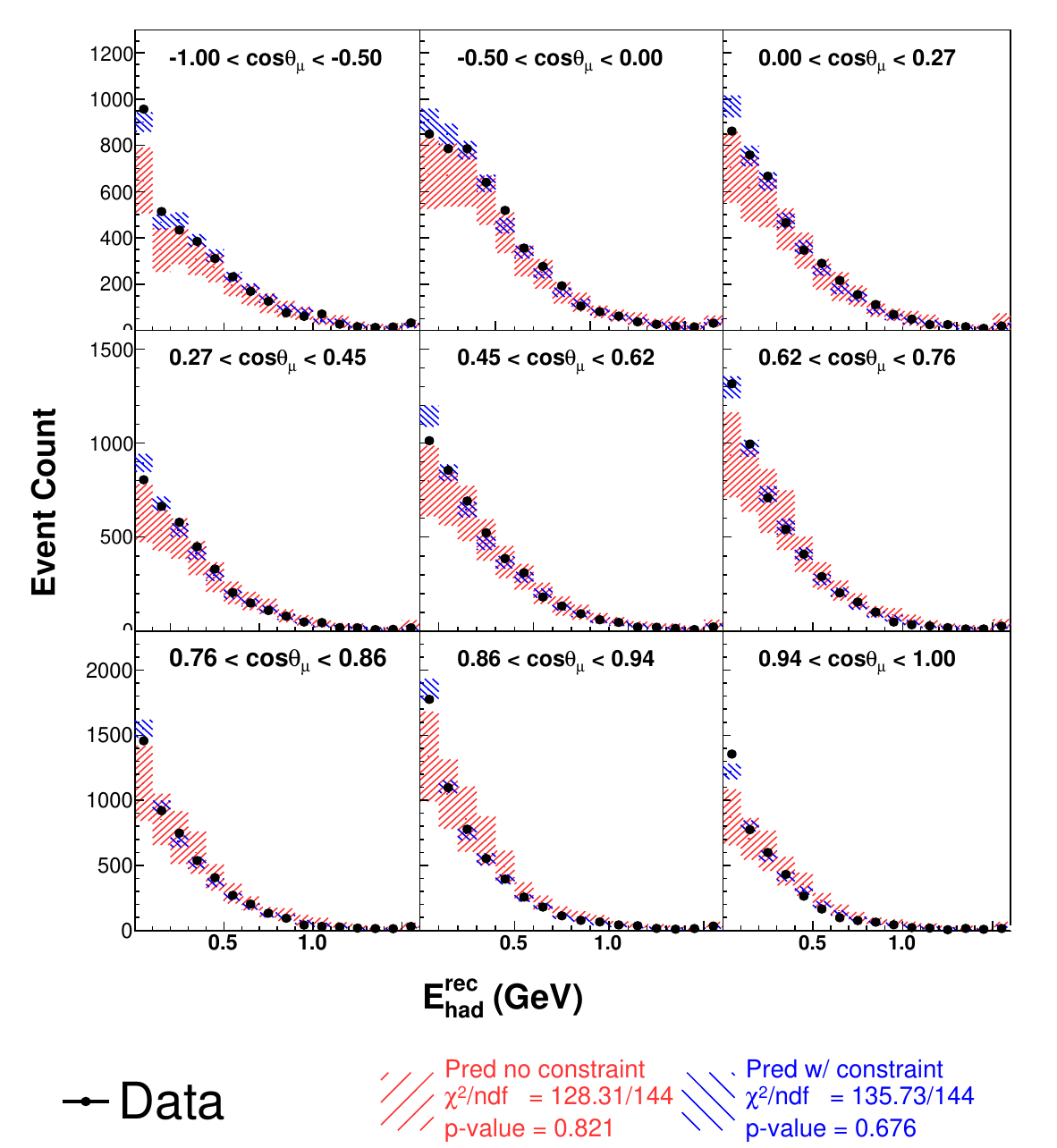}
     \put(-331.9, 175){\color{white}{\rule{0.46cm}{0.3cm}}}
     \put(-331.9, 280){\color{white}{\rule{0.46cm}{0.3cm}}}
     \put(-300,388){\normalsize{MicroBooNE $6.4 \times 10^{20}$ POT}}
     \put(-360,184){\rotatebox{90}{\color{white}{\rule{3cm}{0.8cm}}}}
     \put(-350,200){\rotatebox{90}{\normalsize{Event Count}}}
     \put(-245,36){\color{white}{\rule{4cm}{0.8cm}}}
     \put(-190,50){\normalsize{$E_{\mathrm{had}}^{\mathrm{rec}} (\mathrm{GeV})$}}
     \put(-262,  378){\tiny{rec}}
     \put(-162.5,378){\tiny{rec}}
     \put( -61,  378){\tiny{rec}}
     \put(-264.7,275.0){\tiny{rec}}
     \put(-165.5,275.0){\tiny{rec}}
     \put( -61.3,275.0){\tiny{rec}}
     \put(-260.5,171.2){\tiny{rec}}
     \put(-165.5,171.2){\tiny{rec}}
     \put( -61.3,171.2){\tiny{rec}}
     \caption{Distribution of data and prediction over the 2D reconstructed binning of $\{ E_{\mathrm{had}}^{\mathrm{rec}}, \cos(\theta^{\mathrm{rec}}_{\mu}) \}$ for fully contained events (partially contained event distributions are shown in the supplemental material~\protect{\cite{suppl}}).  The \texttt{MicroBooNE model} prediction, including before (red) and after (blue) applying the measurement of the data distribution over $\{ P^{\mathrm{rec}}_{\mu}, \cos(\theta^{\mathrm{rec}}_{\mu}) \}$ as a constraint, is compared to data. }
    \label{fig:gof}
\end{figure}

Next, a GoF test is performed over the 2D $\{ E_{\mathrm{had}}^{\mathrm{rec}}, \cos(\theta^{\mathrm{rec}}_{\mu}) \}$ distribution, shown in Fig.~\ref{fig:gof}, and is constrained by the muon kinematics measurement using the conditional constraint formalism~\cite{cond_cov}, described in more detail in the supplemental material~\cite{suppl}. It demonstrates a $\chi^{2}/\textrm{ndf}$ of $136/144$ after applying the constraint, again indicating that the model describes the relationship between $\{ P^{\mathrm{rec}}_{\mu}, \cos(\theta^{\mathrm{rec}}_{\mu}) \}$ and $\{ E_{\mathrm{had}}^{\mathrm{rec}} , \cos(\theta^{\mathrm{rec}}_{\mu}) \}$ in data within uncertainties.  The constraint highly suppresses the common uncertainties between these distributions, causing the posterior prediction to have much lower uncertainties and leading to a more stringent examination of the model.  Through the demonstration of accurate muon kinematics modeling, combined with accurate modeling of $E_{\mathrm{had}}^{\mathrm{rec}}$ in relation to $\{ P^{\mathrm{rec}}_{\mu}, \cos(\theta^{\mathrm{rec}}_{\mu}) \}$, the GoF tests validate the modeling of the missing hadronic energy to describe the data within uncertainties.  This builds confidence that the use of the \texttt{MicroBooNE model} in this analysis does not introduce bias beyond the quoted uncertainties.

To help demonstrate the sensitivity of this data-driven model validation approach, a series of fake data studies are performed.  One fake dataset is generated using the \texttt{NuWro} model prediction, and others are generated by varying the reconstructed proton energy of events simulated using the \texttt{MicroBooNE model}, effectively modifying the missing energy in accordance with energy conservation and its relationship to the observable lepton and hadronic energies.  Each fake data study compares the sensitivity in the GoF test $\chi^{2}/\textrm{ndf}$ to the corresponding level of bias in the unfolded $d^{2}\sigma(E_{\nu})/d\cos(\theta_{\mu})dP_{\mu}$ measurement when compared to the underlying truth. The unfolding to neutrino energy was used here as it has a significant potential for model dependence, making it a conservative test study for an unfolding to $E_{\mathrm{vis}}$.  In all cases, the GoF tests demonstrate higher sensitivity to mismodeling than the extracted measurements show bias from unfolding to the three-dimensional binning used in the analysis.  In each of these cases the bias on the unfolded measurement is within the total model uncertainties, and furthermore is either within the cross section and statistical uncertainties or is detected by the GoF test.  See the supplemental material~\cite{suppl} for details on the three-dimensional fake data studies.

\begin{figure*}[hbt!]
     \centering
     \includegraphics[clip, trim={0.5cm 0.15cm 0.5cm 0.65cm}, width=0.98\textwidth]{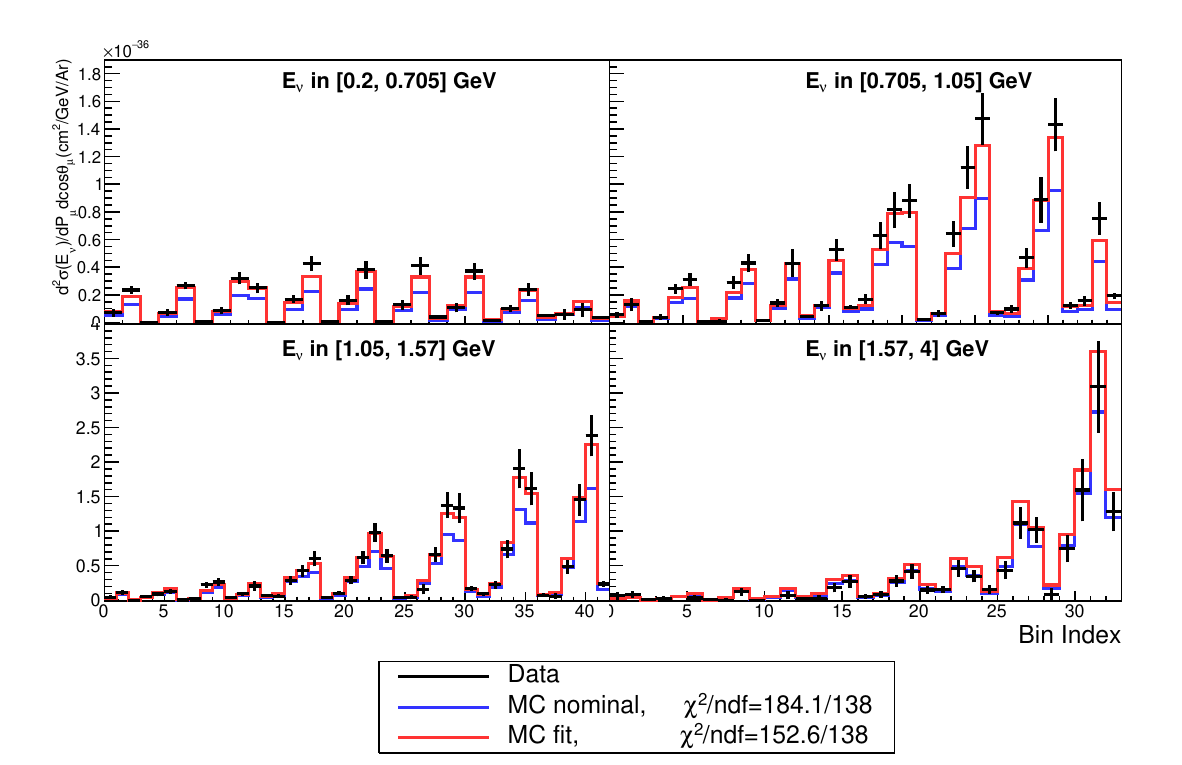}
     \put(-332,235){ MicroBooNE $6.4 \times 10^{20}$ POT}
     \put(-360.5, 141){\color{white}{\rule{0.3cm}{0.3cm}}}
     \put(-185.0, 42.5){\color{white}{\rule{0.3cm}{0.3cm}}}
     \put(-333.6, 144.7){\color{gray}{\rule{0.04cm}{2.25cm}}} 
     \put(-315.5, 144.7){\color{gray}{\rule{0.04cm}{2.25cm}}}
     \put(-291.3, 144.7){\color{gray}{\rule{0.04cm}{2.25cm}}}
     \put(-273.2, 144.7){\color{gray}{\rule{0.04cm}{2.25cm}}}
     \put(-255.0, 144.7){\color{gray}{\rule{0.04cm}{2.25cm}}}
     \put(-236.8, 144.7){\color{gray}{\rule{0.04cm}{2.25cm}}}
     \put(-218.7, 144.7){\color{gray}{\rule{0.04cm}{2.25cm}}}
     \put(-200.3, 144.7){\color{gray}{\rule{0.04cm}{2.25cm}}}
     \put(-344.0, 200){\footnotesize \boldmath{$\theta_{0}$}}
     \put(-327.7, 200){\footnotesize \boldmath{$\theta_{1}$}}
     \put(-306.0, 200){\footnotesize \boldmath{$\theta_{2}$}}
     \put(-285.9, 200){\footnotesize \boldmath{$\theta_{3}$}}
     \put(-267.0, 200){\footnotesize \boldmath{$\theta_{4}$}}
     \put(-248.0, 200){\footnotesize \boldmath{$\theta_{5}$}}
     \put(-230.0, 200){\footnotesize \boldmath{$\theta_{6}$}}
     \put(-212.2, 200){\footnotesize \boldmath{$\theta_{7}$}}
     \put(-194.2, 200){\footnotesize \boldmath{$\theta_{8}$}}
     \put(-167.7, 144.7){\color{gray}{\rule{0.04cm}{2.25cm}}} 
     \put(-148.2, 144.7){\color{gray}{\rule{0.04cm}{2.25cm}}}
     \put(-128.5, 144.7){\color{gray}{\rule{0.04cm}{2.25cm}}}
     \put(-113.7, 144.7){\color{gray}{\rule{0.04cm}{2.25cm}}}
     \put(-98.7,  144.7){\color{gray}{\rule{0.04cm}{2.25cm}}}
     \put(-74.5,  144.7){\color{gray}{\rule{0.04cm}{2.25cm}}}
     \put(-50.2,  144.7){\color{gray}{\rule{0.04cm}{2.25cm}}}
     \put(-25.5,  144.7){\color{gray}{\rule{0.04cm}{2.25cm}}}
     \put(-177.0, 200){\footnotesize \boldmath{$\theta_{0}$}}
     \put(-160.2, 200){\footnotesize \boldmath{$\theta_{1}$}}
     \put(-140.2, 200){\footnotesize \boldmath{$\theta_{2}$}}
     \put(-123.0, 200){\footnotesize \boldmath{$\theta_{3}$}}
     \put(-108.8, 200){\footnotesize \boldmath{$\theta_{4}$}}
     \put(-90.0,  200){\footnotesize \boldmath{$\theta_{5}$}}
     \put(-71.0,  200){\footnotesize \boldmath{$\theta_{6}$}}
     \put(-45.2,  200){\footnotesize \boldmath{$\theta_{7}$}}
     \put(-20.8,  200){\footnotesize \boldmath{$\theta_{8}$}}
     \put(-340.0, 51.5){\color{gray}{\rule{0.04cm}{2.25cm}}} 
     \put(-323.5, 51.5){\color{gray}{\rule{0.04cm}{2.25cm}}}
     \put(-307.5, 51.5){\color{gray}{\rule{0.04cm}{2.25cm}}}
     \put(-295.8, 51.5){\color{gray}{\rule{0.04cm}{2.25cm}}}
     \put(-275.2, 51.5){\color{gray}{\rule{0.04cm}{2.25cm}}}
     \put(-250.7, 51.5){\color{gray}{\rule{0.04cm}{2.25cm}}}
     \put(-226.7, 51.5){\color{gray}{\rule{0.04cm}{2.25cm}}}
     \put(-202.3, 51.5){\color{gray}{\rule{0.04cm}{2.25cm}}}
     \put(-348.0, 108){\footnotesize \boldmath{$\theta_{0}$}}
     \put(-335.0, 108){\footnotesize \boldmath{$\theta_{1}$}}
     \put(-319.0, 108){\footnotesize \boldmath{$\theta_{2}$}}
     \put(-304.8, 108){\footnotesize \boldmath{$\theta_{3}$}}
     \put(-288.0, 108){\footnotesize \boldmath{$\theta_{4}$}}
     \put(-265.0, 108){\footnotesize \boldmath{$\theta_{5}$}}
     \put(-240.5, 108){\footnotesize \boldmath{$\theta_{6}$}}
     \put(-216.5, 108){\footnotesize \boldmath{$\theta_{7}$}}
     \put(-197.8, 108){\footnotesize \boldmath{$\theta_{8}$}}
     \put(-167.0, 51.5){\color{gray}{\rule{0.04cm}{2.25cm}}} 
     \put(-146.5, 51.5){\color{gray}{\rule{0.04cm}{2.25cm}}}
     \put(-130.4, 51.5){\color{gray}{\rule{0.04cm}{2.25cm}}}
     \put(-115.0, 51.5){\color{gray}{\rule{0.04cm}{2.25cm}}}
     \put(-94.0,  51.5){\color{gray}{\rule{0.04cm}{2.25cm}}}
     \put(-72.9,  51.5){\color{gray}{\rule{0.04cm}{2.25cm}}}
     \put(-57.6,  51.5){\color{gray}{\rule{0.04cm}{2.25cm}}}
     \put(-36.8,  51.5){\color{gray}{\rule{0.04cm}{2.25cm}}}
     \put(-175.8, 108){\footnotesize \boldmath{$\theta_{0}$}}
     \put(-159.0, 108){\footnotesize \boldmath{$\theta_{1}$}}
     \put(-140.0, 108){\footnotesize \boldmath{$\theta_{2}$}}
     \put(-124.5, 108){\footnotesize \boldmath{$\theta_{3}$}}
     \put(-107.2, 108){\footnotesize \boldmath{$\theta_{4}$}}
     \put(-85.3,  108){\footnotesize \boldmath{$\theta_{5}$}}
     \put(-66.5,  108){\footnotesize \boldmath{$\theta_{6}$}}
     \put(-49.2,  108){\footnotesize \boldmath{$\theta_{7}$}}
     \put(-30.0,  108){\footnotesize \boldmath{$\theta_{8}$}}
     \caption{ Unfolded measurement of the inclusive $\nu_{\mu}$ CC double-differential cross section on argon as a function of neutrino energy and nominal and tuned \texttt{MicroBooNE} model predictions are shown across the $\{P_{\mu}, \cos(\theta_{\mu})\}$ binning within each $E_{\nu}$ slice. Angle slices are labeled and separated by gray lines, with bin edges $\{ -1, -0.5, 0, 0.27, 0.45, 0.62, 0.76, 0.86, 0.94, 1 \}$. A complete description of the phase space location of each analysis bin is given in the supplemental material~\protect{\cite{suppl}}. Uncertainties on the measurement as well as the \texttt{MicroBooNE} model are combined as error bars on the measurement and both included in the $\chi^{2}$ computation.}
    \label{fig:xs_enu}
\end{figure*}

\begin{figure*}[hbt!]
     \centering
     \includegraphics[clip, trim={0.5cm 0.15cm 0.5cm 0.65cm}, width=0.98\textwidth]{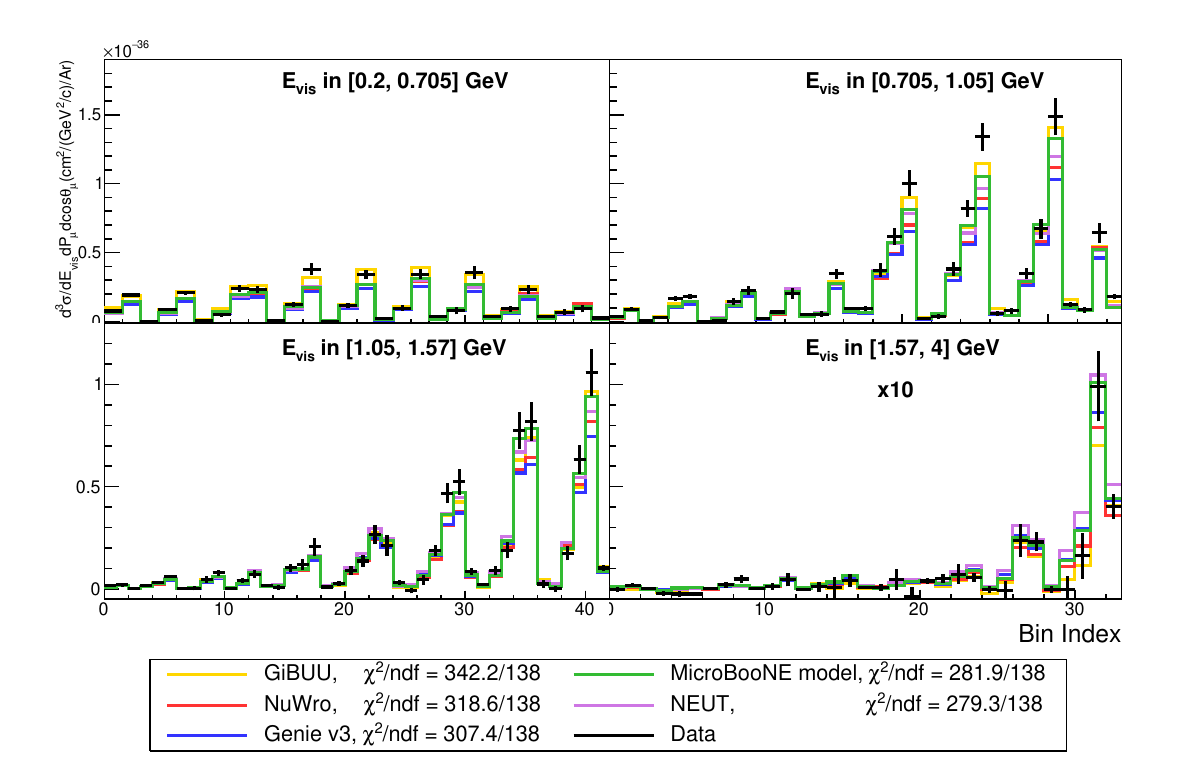}
     \put(-330,235){MicroBooNE $6.4 \times 10^{20}$ POT}
     \put(-360.5, 141){\color{white}{\rule{0.3cm}{0.3cm}}}
     \put(-185.0, 42.5){\color{white}{\rule{0.3cm}{0.3cm}}}
     \put(-333.6, 144.7){\color{gray}{\rule{0.04cm}{2.25cm}}} 
     \put(-315.5, 144.7){\color{gray}{\rule{0.04cm}{2.25cm}}}
     \put(-291.3, 144.7){\color{gray}{\rule{0.04cm}{2.25cm}}}
     \put(-273.2, 144.7){\color{gray}{\rule{0.04cm}{2.25cm}}}
     \put(-255.0, 144.7){\color{gray}{\rule{0.04cm}{2.25cm}}}
     \put(-236.8, 144.7){\color{gray}{\rule{0.04cm}{2.25cm}}}
     \put(-218.7, 144.7){\color{gray}{\rule{0.04cm}{2.25cm}}}
     \put(-200.3, 144.7){\color{gray}{\rule{0.04cm}{2.25cm}}}
     \put(-344.0, 200){\footnotesize \boldmath{$\theta_{0}$}}
     \put(-327.7, 200){\footnotesize \boldmath{$\theta_{1}$}}
     \put(-306.0, 200){\footnotesize \boldmath{$\theta_{2}$}}
     \put(-285.9, 200){\footnotesize \boldmath{$\theta_{3}$}}
     \put(-267.0, 200){\footnotesize \boldmath{$\theta_{4}$}}
     \put(-248.0, 200){\footnotesize \boldmath{$\theta_{5}$}}
     \put(-230.0, 200){\footnotesize \boldmath{$\theta_{6}$}}
     \put(-212.2, 200){\footnotesize \boldmath{$\theta_{7}$}}
     \put(-194.2, 200){\footnotesize \boldmath{$\theta_{8}$}}
     \put(-167.7, 144.7){\color{gray}{\rule{0.04cm}{2.25cm}}} 
     \put(-148.2, 144.7){\color{gray}{\rule{0.04cm}{2.25cm}}}
     \put(-128.5, 144.7){\color{gray}{\rule{0.04cm}{2.25cm}}}
     \put(-113.7, 144.7){\color{gray}{\rule{0.04cm}{2.25cm}}}
     \put(-98.7,  144.7){\color{gray}{\rule{0.04cm}{2.25cm}}}
     \put(-74.5,  144.7){\color{gray}{\rule{0.04cm}{2.25cm}}}
     \put(-50.2,  144.7){\color{gray}{\rule{0.04cm}{2.25cm}}}
     \put(-25.5,  144.7){\color{gray}{\rule{0.04cm}{2.25cm}}}
     \put(-177.0, 200){\footnotesize \boldmath{$\theta_{0}$}}
     \put(-160.2, 200){\footnotesize \boldmath{$\theta_{1}$}}
     \put(-140.2, 200){\footnotesize \boldmath{$\theta_{2}$}}
     \put(-123.0, 200){\footnotesize \boldmath{$\theta_{3}$}}
     \put(-108.8, 200){\footnotesize \boldmath{$\theta_{4}$}}
     \put(-90.0,  200){\footnotesize \boldmath{$\theta_{5}$}}
     \put(-67.0,  200){\footnotesize \boldmath{$\theta_{6}$}}
     \put(-45.2,  200){\footnotesize \boldmath{$\theta_{7}$}}
     \put(-20.8,  200){\footnotesize \boldmath{$\theta_{8}$}}
     \put(-340.0, 51.5){\color{gray}{\rule{0.04cm}{2.25cm}}} 
     \put(-323.5, 51.5){\color{gray}{\rule{0.04cm}{2.25cm}}}
     \put(-307.5, 51.5){\color{gray}{\rule{0.04cm}{2.25cm}}}
     \put(-295.8, 51.5){\color{gray}{\rule{0.04cm}{2.25cm}}}
     \put(-275.2, 51.5){\color{gray}{\rule{0.04cm}{2.25cm}}}
     \put(-250.7, 51.5){\color{gray}{\rule{0.04cm}{2.25cm}}}
     \put(-226.7, 51.5){\color{gray}{\rule{0.04cm}{2.25cm}}}
     \put(-202.3, 51.5){\color{gray}{\rule{0.04cm}{2.25cm}}}
     \put(-348.0, 108){\footnotesize \boldmath{$\theta_{0}$}}
     \put(-335.0, 108){\footnotesize \boldmath{$\theta_{1}$}}
     \put(-319.0, 108){\footnotesize \boldmath{$\theta_{2}$}}
     \put(-304.8, 108){\footnotesize \boldmath{$\theta_{3}$}}
     \put(-288.0, 108){\footnotesize \boldmath{$\theta_{4}$}}
     \put(-265.0, 108){\footnotesize \boldmath{$\theta_{5}$}}
     \put(-240.5, 108){\footnotesize \boldmath{$\theta_{6}$}}
     \put(-222.0, 108){\footnotesize \boldmath{$\theta_{7}$}}
     \put(-197.8, 108){\footnotesize \boldmath{$\theta_{8}$}}
     \put(-167.0, 51.5){\color{gray}{\rule{0.04cm}{2.25cm}}} 
     \put(-146.5, 51.5){\color{gray}{\rule{0.04cm}{2.25cm}}}
     \put(-130.4, 51.5){\color{gray}{\rule{0.04cm}{2.25cm}}}
     \put(-115.0, 51.5){\color{gray}{\rule{0.04cm}{2.25cm}}}
     \put(-94.0,  51.5){\color{gray}{\rule{0.04cm}{2.25cm}}}
     \put(-72.9,  51.5){\color{gray}{\rule{0.04cm}{2.25cm}}}
     \put(-57.6,  51.5){\color{gray}{\rule{0.04cm}{2.25cm}}}
     \put(-36.8,  51.5){\color{gray}{\rule{0.04cm}{2.25cm}}}
     \put(-175.8, 108){\footnotesize \boldmath{$\theta_{0}$}}
     \put(-159.0, 108){\footnotesize \boldmath{$\theta_{1}$}}
     \put(-140.0, 108){\footnotesize \boldmath{$\theta_{2}$}}
     \put(-124.5, 108){\footnotesize \boldmath{$\theta_{3}$}}
     \put(-107.2, 108){\footnotesize \boldmath{$\theta_{4}$}}
     \put(-85.3,  108){\footnotesize \boldmath{$\theta_{5}$}}
     \put(-66.5,  108){\footnotesize \boldmath{$\theta_{6}$}}
     \put(-49.2,  108){\footnotesize \boldmath{$\theta_{7}$}}
     \put(-30.0,  108){\footnotesize \boldmath{$\theta_{8}$}}
     \caption{Unfolded measurement of the inclusive $\nu_{\mu}$ CC triple-differential cross section on argon and various model predictions are shown across the ${P_{\mu}, \cos(\theta_{\mu})}$ binning within each $E_{\mathrm{vis}}$ slice. Angle slices are labeled and separated by gray lines, with bin edges $\{ -1, -0.5, 0, 0.27, 0.45, 0.62, 0.76, 0.86, 0.94, 1 \}$. A complete description of the phase space location of each analysis bin and a comparison to the \texttt{MicroBooNE model} is given in the supplemental material~\protect{\cite{suppl}}. The highest $E_{\mathrm{vis}}$ slice is magnified by a factor of 10 for visibility. }
    \label{fig:xs_evis}
\end{figure*}

This work extracts the triple-differential cross section $d^3\sigma/dE_{\mathrm{vis}}d\cos(\theta_{\mu})dP_{\mu}$ and the three-dimensional cross section $d^2\sigma(E_{\nu})/d\cos(\theta_{\mu})dP_{\mu}$ using the Wiener-SVD unfolding technique~\cite{Tang:2017rob}.  In each case, a regularization term is constructed from matrices that compute the third derivative of the unfolded distribution with respect to each of $E_{\mathrm{vis}}$ (or $E_{\nu}$), $\cos(\theta_{\mu})$, and $P_{\mu}$ by taking differences of nearby bins, and are further combined in quadrature.  The covariance matrix includes statistical uncertainties, computed using the combined Neyman-Pearson method~\cite{Ji:2019yca}, as well as systematic uncertainties for signal and background events.  The bias introduced in unfolding and regularization is captured in an additional smearing matrix $A_{C}$ that is applied to every theoretical prediction reported in this work and included in the data release in the supplemental material~\cite{suppl}. The unfolded cross section consists of 138 bins spanning 4 $E_{\mathrm{vis}}$ (or $E_{\nu}$) slices, 9 $\cos(\theta_{\mu})$ slices, and 3--6 $P_{\mu}$ bins within each slice based on the detector resolution and statistics available.


\begin{figure*}[hbt!]
     \centering
     \includegraphics[clip,trim={0.8cm 0.1cm 1cm 0.9cm}, width=0.95\textwidth]{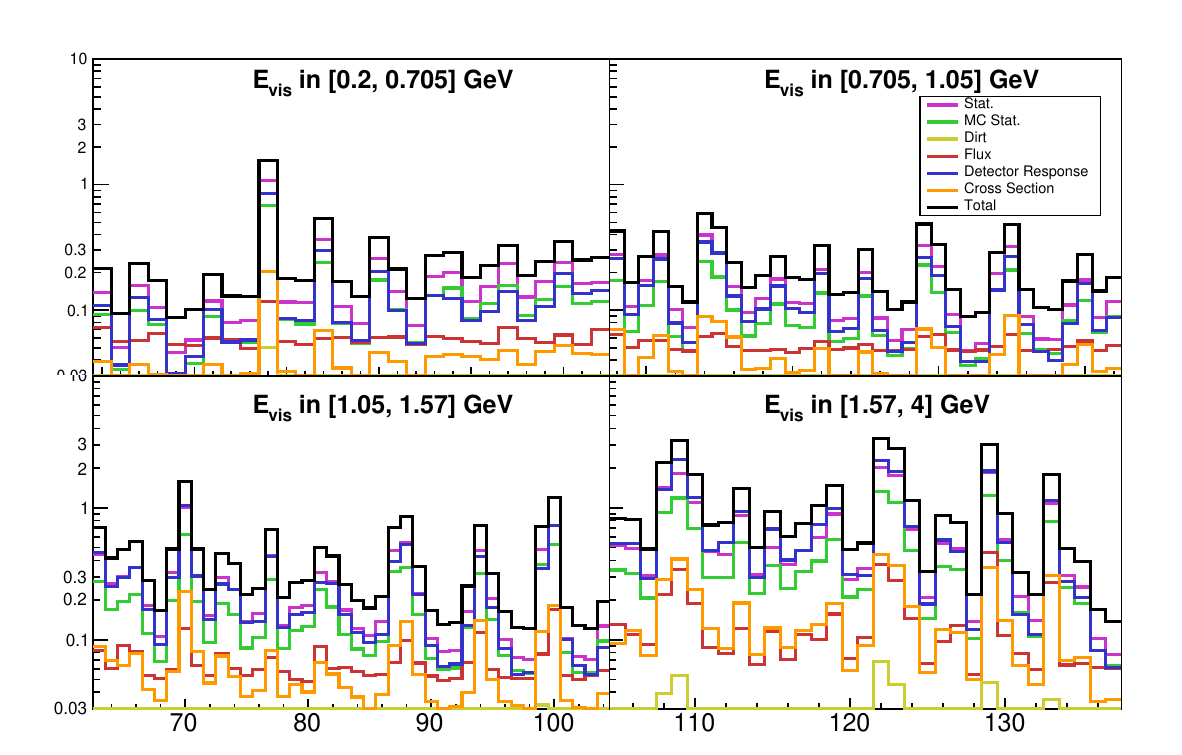}
     \put(-363.0, 124.0){\color{white}{\rule{0.46cm}{0.2cm}}}
     \put(-331.0, 126.8){\color{gray}{\rule{0.04cm}{3.2cm}}} 
     \put(-312.4, 126.8){\color{gray}{\rule{0.04cm}{3.2cm}}}
     \put(-287.2, 126.8){\color{gray}{\rule{0.04cm}{3.2cm}}}
     \put(-268.6, 126.8){\color{gray}{\rule{0.04cm}{3.2cm}}}
     \put(-249.8, 126.8){\color{gray}{\rule{0.04cm}{3.2cm}}}
     \put(-230.9, 126.8){\color{gray}{\rule{0.04cm}{3.2cm}}}
     \put(-212.1, 126.8){\color{gray}{\rule{0.04cm}{3.2cm}}}
     \put(-193.2, 126.8){\color{gray}{\rule{0.04cm}{3.2cm}}}
     \put(-343.0, 210){\footnotesize \boldmath{$\theta_{0}$}}
     \put(-324.5, 210){\footnotesize \boldmath{$\theta_{1}$}}
     \put(-302.0, 210){\footnotesize \boldmath{$\theta_{2}$}}
     \put(-280.9, 210){\footnotesize \boldmath{$\theta_{3}$}}
     \put(-263.0, 210){\footnotesize \boldmath{$\theta_{4}$}}
     \put(-244.0, 210){\footnotesize \boldmath{$\theta_{5}$}}
     \put(-224.0, 210){\footnotesize \boldmath{$\theta_{6}$}}
     \put(-205.2, 210){\footnotesize \boldmath{$\theta_{7}$}}
     \put(-188.2, 210){\footnotesize \boldmath{$\theta_{8}$}}
     \put(-159.6,  126.8){\color{gray}{\rule{0.04cm}{3.2cm}}} 
     \put(-139.8,  126.8){\color{gray}{\rule{0.04cm}{3.2cm}}}
     \put(-120.0,  126.8){\color{gray}{\rule{0.04cm}{3.2cm}}}
     \put(-105.0,  126.8){\color{gray}{\rule{0.04cm}{3.2cm}}}
     \put(-90.2,   126.8){\color{gray}{\rule{0.04cm}{3.2cm}}}
     \put(-65.3,   126.8){\color{gray}{\rule{0.04cm}{1.9cm}}}
     \put(-40.6,   126.8){\color{gray}{\rule{0.04cm}{1.9cm}}}
     \put(-15.8,   126.8){\color{gray}{\rule{0.04cm}{1.9cm}}}
     \put(-172.0,  210){\footnotesize \boldmath{$\theta_{0}$}}
     \put(-154.5,  210){\footnotesize \boldmath{$\theta_{1}$}}
     \put(-134.2,  210){\footnotesize \boldmath{$\theta_{2}$}}
     \put(-116.2,  210){\footnotesize \boldmath{$\theta_{3}$}}
     \put(-100.9,  210){\footnotesize \boldmath{$\theta_{4}$}}
     \put(-83.0,   210){\footnotesize \boldmath{$\theta_{5}$}}
     \put(-337.4,  13.5){\color{gray}{\rule{0.04cm}{3.35cm}}} 
     \put(-320.8,  13.5){\color{gray}{\rule{0.04cm}{3.35cm}}}
     \put(-304.1,  13.5){\color{gray}{\rule{0.04cm}{3.35cm}}}
     \put(-291.4,  13.5){\color{gray}{\rule{0.04cm}{3.35cm}}}
     \put(-270.5,  13.5){\color{gray}{\rule{0.04cm}{3.35cm}}}
     \put(-245.4,  13.5){\color{gray}{\rule{0.04cm}{3.35cm}}}
     \put(-220.4,  13.5){\color{gray}{\rule{0.04cm}{3.35cm}}}
     \put(-195.3,  13.5){\color{gray}{\rule{0.04cm}{3.35cm}}}
     \put(-345.5,  103){\footnotesize \boldmath{$\theta_{0}$}}
     \put(-332.0,  103){\footnotesize \boldmath{$\theta_{1}$}}
     \put(-315.0,  103){\footnotesize \boldmath{$\theta_{2}$}}
     \put(-299.7,  103){\footnotesize \boldmath{$\theta_{3}$}}
     \put(-284.0,  103){\footnotesize \boldmath{$\theta_{4}$}}
     \put(-262.0,  103){\footnotesize \boldmath{$\theta_{5}$}}
     \put(-235.8,  103){\footnotesize \boldmath{$\theta_{6}$}}
     \put(-211.8,  103){\footnotesize \boldmath{$\theta_{7}$}}
     \put(-189.0,  103){\footnotesize \boldmath{$\theta_{8}$}}
     \put(-158.8,  13.5){\color{gray}{\rule{0.04cm}{3.35cm}}} 
     \put(-137.7,  13.5){\color{gray}{\rule{0.04cm}{3.35cm}}}
     \put(-122.0,  13.5){\color{gray}{\rule{0.04cm}{3.35cm}}}
     \put(-106.3,  13.5){\color{gray}{\rule{0.04cm}{3.35cm}}}
     \put(-85.1,   13.5){\color{gray}{\rule{0.04cm}{3.35cm}}}
     \put(-64.2,   13.5){\color{gray}{\rule{0.04cm}{3.35cm}}}
     \put(-48.4,   13.5){\color{gray}{\rule{0.04cm}{3.35cm}}}
     \put(-27.4,   13.5){\color{gray}{\rule{0.04cm}{3.35cm}}}
     \put(-169.5,  103){\footnotesize \boldmath{$\theta_{0}$}}
     \put(-146.5,  103){\footnotesize \boldmath{$\theta_{1}$}}
     \put(-133.0,  103){\footnotesize \boldmath{$\theta_{2}$}}
     \put(-117.5,  103){\footnotesize \boldmath{$\theta_{3}$}}
     \put(-99.0,   103){\footnotesize \boldmath{$\theta_{4}$}}
     \put(-76.0,   103){\footnotesize \boldmath{$\theta_{5}$}}
     \put(-59.5,   103){\footnotesize \boldmath{$\theta_{6}$}}
     \put(-40.2,   103){\footnotesize \boldmath{$\theta_{7}$}}
     \put(-16.5,   103){\footnotesize \boldmath{$\theta_{8}$}}
     \put(-340,236){\normalsize{MicroBooNE $6.4 \times 10^{20}$ POT}}
     \put(-378,125){\rotatebox{90}{\normalsize Fractional Uncertainty}}
     \put(-70,-5){\normalsize {Bin Index}}
     \caption{The ratio of the diagonal uncertainty to the measurement value in each bin is shown, including breakdown by source of the uncertainty.  Angle slices are labeled and separated by gray lines, with bin edges $\{ -1, -0.5, 0, 0.27, 0.45, 0.62, 0.76, 0.86, 0.94, 1 \}$. A complete description of the phase space location of each analysis bin is given in the supplemental material.  }
    \label{fig:uncertainties}
\end{figure*}

Using this unfolding approach, we demonstrate the utility of data-driven model validation in enabling robust model tuning—a critical ingredient for precision neutrino oscillation measurements. The unfolded $d^{2}\sigma(E_{\nu})/d\cos(\theta_{\mu})dP_{\mu}$ cross section is compared to a \texttt{GENIE} model prediction that approximates the MicroBooNE model prediction, allowing the same four model parameters to vary within their prior uncertainties as used in the referenced tune~\cite{MicroBooNE:2021ccs}. Note that these parameters were originally selected to cover the quasi-elastic-dominated measurement used to tune the \texttt{MicroBooNE model} and may not fully cover the inclusive phase space in this measurement, although this does not preclude their use in this tuning demonstration. While an unfolding on neutrino energy is model-dependent, it is fundamentally implied in any near detector analysis performed in oscillation measurements at long-baseline experiments. The sets of validation studies shown in this paper offer a robust approach to validate such measurements. This model tuning is intended as a proof of concept, and is not used to extract model parameter best-fit values for any physics use.

The fit attempt, shown in Fig.~\ref{fig:xs_enu}, is achieved following the extensive supporting validations of the MicroBooNE model in comparison to the data over relevant kinematic distributions. The fit $\chi^{2}/\textrm{ndf}$ improves from $184.1/138$ to $152.6/138$, and is computed using a combined covariance matrix of the data and model uncertainties. More details on the fit procedure, including parameter best-fit values provided for completeness, are provided in the supplemental material~\cite{suppl}. The most notable change in parameter values is an increase in $M_{A}$, which is consistent with the higher overall normalization of the cross section across the phase space in both data and the post-fit model prediction. Visually, the fit model prediction shows overall better agreement with the data across the 3D measurement space. This is particularly notable given the 3D phase space, including the difficult-to-reconstruct neutrino energy. We note an absence of any visible Peele's Pertinent Puzzle~\cite{fruhwirth2012peelle} issues such as in the normalization agreement in the fit, which has been observed in the model fitting of previous neutrino-nuclei cross-section data~\cite{chakrani2023parametrized}. The model validation approach underpinning this fit demonstration has wide applications including supporting neutrino cross section and oscillation measurements.


Following the same unfolding approach, but using $E_{\mathrm{vis}}$ instead of $E_{\nu}$, we report the unfolded triple-differential cross section $d^{3}\sigma/dE_{\mathrm{vis}}d\cos(\theta_{\mu})dP_{\mu}$, shown in Fig.~\ref{fig:xs_evis}. The uncertainty contribution from each source as a function of the measured cross section is shown in Fig.~\ref{fig:uncertainties}. We report a total cross section of $0.32 \pm 0.017 \times 10^{-36}\,\mathrm{cm}^{2} \mathrm{Ar}^{-1}$, computed by collapsing the triple-differential measurement $d^{3}\sigma/dE_{\mathrm{vis}}d\cos(\theta_{\mu})dP_{\mu}$ over $E_{\nu}$, $\cos(\theta_{\mu})$, and $P_{\mu}$.
Also included are comparisons with model predictions for the $\texttt{MicroBooNE}$ model, \texttt{GENIE v3.0.6 G18\_10a\_02\_11a}~\cite{GENIE:2021npt} (\texttt{GENIE v3 untuned}), \texttt{GiBUU 2025 patch 1}~\cite{Buss:2011mx} (\texttt{GiBUU}), \texttt{NEUT 5.4.0.1}~\cite{Hayato:2009zz} (\texttt{NEUT}), and \texttt{NuWro 21.09.02}~\cite{Golan:2012rfa} (\texttt{NuWro}), generated using the \texttt{NUISANCE} package~\cite{Stowell:2016jfr}.  A comparison of the underlying physics models in these event generators can be found in Ref.~\cite{Avanzini:2021qlx}.  The unfolded three-dimensional measurement is found to be in tension with all model CV predictions, with \texttt{NEUT} and the \texttt{MicroBooNE model} showing the best agreement. As shown in the supplemental material~\cite{suppl}, \texttt{GiBUU} yields the best result at describing the data within each individual $E_{\mathrm{vis}}$ slice with $\chi^{2}/\textrm{ndf}$ of $36.7/28$, $58.8/35$, $48.9/42$, and $45.8/33$.  However, it struggles to describe the correlations between $E_{\mathrm{vis}}$ slices, leading to the worst overall agreement with the data, mirroring the trend seen in a MicroBooNE measurement of mesonless $\nu_{\mu}$ CC interactions~\cite{gardiner_ccnp}.  Owing to the improved level of detail available across the three-dimensional phase space, the power of these results in differentiating models is significantly improved compared to the previous single-differential analysis.  In general, comparisons to model predictions must be made using the full measurement and covariance matrix.

Looking forward, this measurement can be enhanced by using the increased statistics of the full BNB dataset, as well as by combining MicroBooNE data from the BNB and the Neutrinos at the Main Injector beamline~\cite{numi} to further increase the statistics, while reducing the flux-related uncertainties.  Furthermore, measurements of the cross section in semi-inclusive and exclusive channels will allow for investigation of the modeling of the hadronic final states. 

In summary, building on a comprehensive model validation—including consistency checks across reconstructed kinematic variables and the mapping between true neutrino energy and visible observables—we report the nominal-flux-averaged differential inclusive $\nu_\mu$ CC cross section on argon $d^{3}\sigma/dE_{\mathrm{vis}} d\cos(\theta_{\mu}) dP_{\mu}$, using an exposure of $6.4 \times 10^{20}$ POT of data from the Booster Neutrino Beam at Fermilab.  Comparisons with model predictions show the best agreement with \texttt{NEUT} and the \texttt{MicroBooNE model}, however, no model is able to describe the measurement within uncertainties across all energy bins.  This work advances the field of cross section physics by providing a triple-differential measurement over a complete three-dimensional kinematic phase space for inclusive $\nu_\mu$ CC scattering on argon.  This allows for a better understanding of neutrino event generator performance across a broad phase space.

\section*{Acknowledgments}
This document was prepared by the MicroBooNE collaboration using the
resources of the Fermi National Accelerator Laboratory (Fermilab), a
U.S. Department of Energy, Office of Science, HEP User Facility.
Fermilab is managed by Fermi Research Alliance, LLC (FRA), acting
under Contract No. DE-AC02-07CH11359.  MicroBooNE is supported by the
following: the U.S. Department of Energy, Office of Science, Offices
of High Energy Physics and Nuclear Physics; the U.S. National Science
Foundation; the Swiss National Science Foundation; the Science and
Technology Facilities Council (STFC), part of the United Kingdom Research 
and Innovation; the Royal Society (United Kingdom); and the UK Research 
and Innovation (UKRI) Future Leaders Fellowship. Additional support for 
the laser calibration system and cosmic ray tagger was provided by the 
Albert Einstein Center for Fundamental Physics, Bern, Switzerland. We 
also acknowledge the contributions of technical and scientific staff 
to the design, construction, and operation of the MicroBooNE detector 
as well as the contributions of past collaborators to the development 
of MicroBooNE analyses, without whom this work would not have been 
possible. For the purpose of open access, the authors have applied 
a Creative Commons Attribution (CC BY) public copyright license to 
any Author Accepted Manuscript version arising from this submission.

\bibliographystyle{1_elsarticle-num} 

\end{document}